\documentclass[aps,prd,amsmath,twocolumn,amssymbaps,showpacs]{revtex4-1}
\usepackage{graphicx}  % needed for figures
\usepackage{dcolumn}   % needed for some tables
\usepackage{bm}        % for math
\usepackage{amssymb}   % for math
\usepackage{amsmath}   % for math
\usepackage{lineno}
\usepackage{draftcopy}
\usepackage{relsize} %  ! da scommentare alla fine
\usepackage{xfrac}    % ! da scommentare alla fine
\usepackage{slashed}

\usepackage{color}
\definecolor{dgreen}{cmyk}{1.,0.,1.,0.2}        % dark green
\definecolor{orange}{cmyk}{0.,0.353,1.,0.}    % orange

\usepackage[bookmarks]{hyperref}

\newcommand{\di}{{\rm d}}

\newcommand{\p}{{\rm p}}

\newcommand{\be}{\begin{equation}}
\newcommand{\ee}{\end{equation}}                                                                               
\newcommand{\bea}{\begin{eqnarray}}
\newcommand{\eea}{\end{eqnarray}}

\begin{document}
\title{The electromagnetic field effects in in-out and in-in formalisms}

\author{Gaoqing Cao}
\affiliation{School of Physics and Astronomy, Sun Yat-sen University, Guangzhou 510275, China.}
\date{\today}

\begin{abstract}
In this work, we compare the effects of electromagnetic (EM) fields on strong coupling systems in two formalisms: in-out and in-in, which are different from each other when a finite electric field is present. The chiral effective Nambu--Jona-Lasinio model is adopted for this study, and two kinds of EM field distributions are considered: pure electric field and parallel EM (PEM) field with equal electric and magnetic components. For both distributions, we find that the results of in-out and in-in formalisms start to diverge when the electric field $(qE)^{1/2}$ is larger than the chiral effective mass $M=[m^2+(\pi^0)^2]^{1/2}$, that is, when the Schwinger pair production mechanism becomes important. Besides, the chiral restorations are stiffer in the in-in formalism, especially the transition shifts to first-order instead of the second-order for the PEM field. The neutral collective modes are also explored accordingly: For the PEM field, more precise calculations show nonmonotonic features of their pole masses due to parity mixing, and then the Goldstone-like mode is found to be noneffective at the end of chiral rotation because of chiral anomaly.
\end{abstract}

%\pacs{ }

\maketitle
\section{Introduction}
In peripheral heavy ion collisions, the strongest electric ($E$) and magnetic ($B$) fields of recent Universe were expected to be produced in the fireball~\cite{Skokov:2009qp,Deng:2012pc,Deng:2014uja,Bloczynski:2012en}, due to the short distance between heavy ions and large Lorentz factors for their relativistic motions, respectively. The magnitude of the magnetic field is $10^{18}-10^{20}~{\rm Gauss}$ or $(0.1-1~{\rm GeV})^2$ in natural unit, which is comparable to the energy scale of QCD: $\Lambda_{\rm QCD}\sim 0.2~{\rm GeV}$ thus would have a non-negligible effect on the ground state. By following such philosophy, numerical simulations were carried out in the first-principle lattice QCD which wouldn't suffer from {\it sign problem} for finite $B$. While confirming the sophisticated {\it magnetic catalysis effect} at low temperature~\cite{Gusynin:1994re,Gusynin:1994xp}, they also discovered an unexpected feature of QCD around the pseudo-critical temperature, the so called "{\it inverse magnetic catalysis effect}~\cite{Bali:2011qj,Bali:2012zg}. There are several proposals for this puzzle~\cite{Bruckmann:2013oba,Fukushima:2012kc,Kojo:2012js,Hattori:2015aki,Chao:2013qpa,Cao:2014uva,Ferrer:2014qka,Mao:2016fha} but it remains an open question. Associated with the shift of ground state, the dispersions of either the neutral or charged collective modes (mesons) will change accordingly. For example, the pole mass responds as the following~\cite{Hidaka:2012mz,Bali:2017ian,Avancini:2016fgq,Wang:2017vtn,Mao:2018dqe,Liu:2018zag,Coppola:2018vkw,Cao:2019res}: it decreases for neutral pion and charged rho mesons but increases for charged pions with not too strong magnetic field.

At large enough temperature when chiral symmetry is approximately restored~\cite{Zhuang:1994dw,Klevansky:1989vi}, macroscopic chiral anomaly phenomena~\cite{Kharzeev:2007jp,Fukushima:2008xe,Kharzeev:2010gd,Son:2004tq,Metlitski:2005pr,Huang:2013iia,Hattori:2016njk} can emerge to an observable level, which is now a very important scientific target of the phase II program of Beam Energy Scan in STAR~\cite{Liao:2014ava,Kharzeev:2015znc,Huang:2015oca}. Among all the circumstances for anomaly, the one with parallel electromagnetic (PEM) field is specific because the background is already parity-violating even without involving any matter. In our previous studies~\cite{Cao:2015cka,Wang:2018gmj}, neutral pseudo-scalar mesons were found to condensate under the PEM field; and P. Copinger and K. Fukushima discovered that the Schwinger pair production (SPP) rate is enhanced by a semilocalized static magnetic field~\cite{Copinger:2016llk}. Furthermore, there is one very important discovery for the chiral anomaly dynamics~\cite{Copinger:2018ftr}: In order to understand the contradiction between massless case and small mass limit for chiral anomaly $\partial_\mu J^{\mu}_5$, one has to adopt the in-in (or real-time) rather than the in-out (or imaginary-time) propagator for the fermions. The in-out propagator was derived by Schwinger in 1951 under the proper-time formalism, while the in-in propagator was obtained by solving the real-time Dirac equation directly. The advantage of the in-in propagator is that it captures the right SPP rate in the presence of electric field, see the detailed discussions in Ref.~\cite{Cohen:2008wz}. Physically, the main difference between in-in and in-out schemes is whether the feedback of SPP has been taken into account or not in electric field~\cite{Fradkin:1991}. So we can identify the {\it unobserved yet} Schwinger mechanism indirectly by comparing the mesonic properties detected in heavy ion collisions with those predicted in these formalisms.

It constitutes the main motivation of our recent work to compare the "order parameters" of chiral symmetry and the associated collective modes in both in-out and in-in formalisms. This would give some indications on the validity of the much simpler and more commonly used in-out formalism for the study of electric field effect. The paper is arranged as follows: In Sec.~\ref{gapeqs}, we derive the "gap equations" universally for both in-out and in-in formalisms, which are then regularized in a "vacuum regularization" like scheme. Sec.~\ref{PFIIF} is devoted to the analytical calculations and regularizations of the polarization functions in the in-in formalism. We illuminate our numerical results for pure electric field and PEM field in Sec.~\ref{Numer} and pay special attention to the instability at the end of chiral rotation. A simple summary is given in Sec.~\ref{summary}.

\section{"Gap equations" in parallel electromagnetic field}\label{gapeqs}
For simplicity and without lose of generality, we adopt one-flavor Nambu--Jona-Lasinio (NJL) model for the study of strong coupling systems in the presence of parallel electromagnetic field. The Lagrangian density is given by the following form~\cite{Nambu:1961fr,Nambu:1961tp,Klevansky:1989vi}:
\begin{eqnarray}\label{LNJL}
{\cal L}_{\rm NJL}=\bar\psi(i\slashed{D}-m_0)\psi+G[(\bar\psi\psi)^2+(\bar\psi i\gamma^5\psi)^2],
\end{eqnarray}
where  the spinor $\psi$ represents the $u$ quark field, $m_0$ is the current quark mass, and $G$ is the four-fermion coupling constant. The PEM field is introduced through the covariant derivative $D_\mu=\partial_\mu+iqA_\mu$ with the vector potential chosen explicitly as $A_\mu=(iEz,0,-Bx,0)$ and the $u$ pquark charge $q=2/3e$. The signs of $E$ and $B$ are not so important for our study, thus we stick to the case with $E,B\geq0$. For small $m_0$, the Lagrangian has approximate $U_L(1)\otimes U_R(1)$ chiral symmetry.

In in-out time formalism (IOF), the system is assumed to be in equilibrium and the ground state can be explored by following the standard procedure~\cite{Cao:2015cka,Wang:2018gmj}. We introduce two auxiliary neutral boson fields via the Hubbard-Stratonavich transformation: $\sigma(x)=-2G\bar\psi(x)\psi(x)$ and ${\pi}^0(x)=-2G\bar\psi(x) i\gamma^5\psi(x)$, then the action can be bosonized by integrating out the quark degrees of freedom as:
\begin{eqnarray}\label{action}
{\cal S}_{\rm NJL}\!=\!\int\!{\di^4x}{\sigma^2\!+\!({\pi^0})^2\over 4G}\!-\!{\rm Tr}\ln\!\big[i{\slashed D}\!-\!m_0\!-\!\sigma\!-\!i\gamma^5{\pi}^0\big].
\end{eqnarray}
In recent work, we only consider homogeneous chiral condensates with the expectation values: $\langle\sigma(x)\rangle\equiv m-m_0$ and $\langle\pi^0(x)\rangle\equiv\pi^0$. Then 
the gap equations can be obtained by minimizing the thermodynamic potential $\Omega\equiv {\cal S}_{\rm NJL}/V_4$ with respect to these order parameters, that is, $\partial\Omega/\partial x=0~(x=m,\pi^0)$, as:
\begin{eqnarray}
{m-m_0\over 2G}-{1\over V_4}\text {Tr}\;{\cal S}(x,x')&=&0,\label{gapm}\\
{\pi^0\over 2G}-{1\over V_4}\text {Tr}\;{\cal S}(x,x')i\gamma^5&=&0.\label{gappi0}
\end{eqnarray}
Here, ${\cal S}(x,x')\equiv-\big[i{\slashed D}\!-\!m_0\!-\!\sigma\!-\!i\gamma^5{\pi}^0\big]^{-1}_{xx'}$ is the operator form of quark propagator, $V_4$ is the four dimensional space-time volume in Euclidean space, and the trace should be taken over the coordinate, Dirac and color spaces. 

Expand the action Eq.(\ref{action}) to quadratic orders of the fluctuation fields: $\hat{\sigma}(x)\equiv\sigma(x)-\langle\sigma\rangle$ and $\hat{\pi}^0(x)\equiv{\pi}^0(x)-\langle\pi^0\rangle$ and transform to energy momentum space, the polarization functions can be generally written in the random-phase approximation (RPA) as
\begin{eqnarray}
\Pi_{\rm MM'}(q)=\int\frac{d^4p}{(2\pi)^4}{\rm Tr}~\hat{S}(q+p)\Gamma_{\rm M}\hat{S}(p)\Gamma_{\rm{M'}}
\end{eqnarray}
with the interaction vertices given by
\begin{eqnarray}
\Gamma_{\hat{\sigma}}=1, ~\Gamma_{\hat{\pi}^0}=i\gamma_5.
\end{eqnarray}
In the matrix form, the polarization function can be represented in the parity-doublet space ass:
\begin{eqnarray}
\Pi(q)=\left(\begin{array}{cc}
\Pi_{\hat{\sigma}\hat{\sigma}}(q) & \Pi_{\hat{\sigma}\hat{\pi}^0}(q) \\
\Pi_{\hat{\pi}^0\hat{\sigma}}(q) & \Pi_{\hat{\pi}^0\hat{\pi}^0}(q)
\end{array} \right).
\end{eqnarray}
Then,  the matrix form of the inverse of the effective mesonic propagator follows directly:
\begin{eqnarray}
\mathcal{G}^{-1}(q)=\frac{1}{2G}+\Pi(q).
\end{eqnarray}
By diagonalizing $\mathcal{G}^{-1} $, we obtain the inverse propagators of the mass eigen modes, which we denote by $"\Sigma"$ and $"\Pi^0"$, as:
\begin{eqnarray}
\mathcal{G}^{-1}_{\Sigma}&=&{1\over 2}\left[\frac{1}{G}+\Pi_{\hat{\sigma}+\hat{\pi}^0}+\sqrt{(\Pi_{\hat{\sigma}-\hat{\pi}^0})^2+4(\Pi_{\hat{\sigma}\hat{\pi}^0})^2}\right],\nonumber\\
\mathcal{G}^{-1}_{\Pi^0}&=&{1\over 2}\left[\frac{1}{G}+\Pi_{\hat{\sigma}+\hat{\pi}^0}-\sqrt{(\Pi_{\hat{\sigma}-\hat{\pi}^0})^2+4(\Pi_{\hat{\sigma}\hat{\pi}^0})^2}\right]
\end{eqnarray}
with $\Pi_{\hat{\sigma}\pm\hat{\pi}^0}\equiv\Pi_{\hat{\sigma}\hat{\sigma}}\pm\Pi_{\hat{\pi}^0\hat{\pi}^0}$.

In in-in formalism (IIF), the forms of the "gap equations" would remain the same as Eqs.\eqref{gapm} and \eqref{gappi0} by following the definitions of the auxiliary boson fields or the Hartree approximation. But there is no self-consistent thermodynamic potential for these "gap equations" when the feedback of Schwinger mechanism is included explicitly, see discussions later. Nevertheless, we can still use RPA to explore the effective collective excitations in IIF by following the same chain diagrams as in IOF~\cite{Klevansky:1989vi}. Eventually, the pole masses of the collective modes can be evaluated by setting $\mathcal{G}^{-1}_{\rm M}(q_4=im_{\rm M},{\bf q}=0)=0$ in IOF and $\mathcal{G}^{-1}_{\rm M}(q_0=m_{\rm M},{\bf q}=0)=0$ in IIF with ${\rm M}=\Sigma, \Pi^0$. 

Now, the difference between IOF and IIF just originates from the different forms of quark propagator adopted in the calculations. The explicit form of the propagators and the consequential derivations have already been given in Ref.~\cite{Wang:2018gmj} for IOF, so we mainly focus on carrying out analytic derivations under the IIF in the following. In IIF, the quark propagator can be written out explicitly in coordinate space as ${\cal S}(x,x')=\Phi(A)\tilde{\cal S}(x-x')$ with the Schwinger phase $\Phi(A)\equiv e^{-iq\int_{x'}^xA_\mu dx^\mu}$ and the effective propagator~\cite{Copinger:2018ftr}
\begin{widetext}
	\begin{eqnarray}
	\tilde{\cal S}(y)
	&=&{-i\,qBqE\over(4\pi)^2}\int_{RT}\di s\exp\left\{-iM^2s+{i\over4}\left[{qB\over\tan(qBs)}(y_1^2+y_2^2)+{qE\over\tanh(qEs)}(y_3^2-y_0^2)\right]\right\}\nonumber\\
	&&\Bigg\{m-i\gamma^5\pi^0-{qB\over2}\Big[\big(\cot(qBs)\gamma^1+\gamma^2\big)y_1+\big(\cot(qBs)\gamma^2-\gamma^1\big)y_2\Big]-{qE\over2}\Big[\big(\coth(qEs)\gamma^3-i\gamma^4\big)y_3\nonumber\\
	&&+\big(\coth(qEs)\gamma^4+i\gamma^3\big)i\,y_0\Big]\Bigg\}\Big[\cot(qBs)\coth(qEs)+i\gamma^5
	+\coth(qEs)\gamma^1\gamma^2+i\cot(qBs)\gamma^4\gamma^3\Big].\label{effprop}
	\end{eqnarray}
Here, we've defined variables $y\equiv x-x'$, $M^2\equiv m^2+(\pi^0)^2$ and the RT integral path
\bea\label{RTI}
\int_{RT}&\equiv& \left(\int_{-\infty-iS_E}^{-iS_E-\epsilon}+\int_{\infty-iS_E/2}^{-\infty-iS_E/2}+\int_{0}^{-\infty}\right)\theta(y_3)+\left(\int_{\infty-iS_E}^{-iS_E+\epsilon}+\int_{0}^{\infty}\right)\theta(-y_3)
\eea	
with the SPP parameter $S_E=\pi/|qE|$. In principle, there should be delta function terms that are from the derivatives $\partial_{x_3}\theta(\pm y_3)$ in the effective propagator Eq.~\eqref{effprop}~\cite{Copinger:2018ftr}. But we neglect them as their total contribution is infinitesimally small in energy  momentum space thus not relevant to our recent study, similar to the discussions in Eq.~\eqref{SMB} and below. At this point, the only difference between the in-out and in-in propagators is the additional integral branches in IIF, that is, the $S_E$ dependent terms on the right-hand side of Eq.\eqref{RTI} which account for the feedback of SPP. 

Take the first branch of the path for $y_3>0$ for example and keep only the $m$ and $y_3$ relevant terms, we can use the following representation of $\theta$ function
$$\theta(y_3)={1\over2\pi i}\int_{-\infty}^\infty\di q_3{e^{i q_3y_3}\over q_3-i\eta},\ \ \eta\rightarrow 0^+$$
to carry out Fourier transformation and find:
\begin{eqnarray}\label{SMB}
&&{1\over2\pi i}\lim_{\eta\rightarrow 0^+}\int\di y_3 e^{-i p_3 y_3}\int_{-\infty}^\infty\di q_3{e^{i q_3y_3}\over q_3-i\eta}\int_{-\infty-iS_E}^{-iS_E-\epsilon}\di s~e^{{i\over4}{qE\over\tanh(qEs)}y_3^2}\left[m-{qE\over2}\big(\coth(qEs)\gamma^3-i\gamma^4\big)y_3\right]\nonumber\\
&=&{1\over2\pi i}\lim_{\eta\rightarrow 0^+}\int_{-\infty}^\infty\di q_3{1\over q_3-i\eta}\int_{-\infty}^{-\epsilon}\di s\,e^{-i{\tanh(qEs)\over qE}(p_3-q_3)^2}\left({-i\over4\pi}{qE\over\tanh(qEs)}\right)^{-{1\over2}}\left[m-\big(\gamma^3-i\,\tanh(qEs)\gamma^4\big)(p_3-q_3)\right]\nonumber\\
&=&\int_{-\infty}^{-\epsilon}\di s\left[\left(m-\big(\gamma^3-i\,\tanh(qEs)\gamma^4\big)p_3\right)G\left(p_3,{\tanh(qEs)\over qE}\right)\left({-i\over4\pi}{qE\over\tanh(qEs)}\right)^{-{1\over2}}-i\big(\gamma^3-i\,\tanh(qEs)\gamma^4\big)\right],
\end{eqnarray}
where the auxiliary function is
$$G\left(p_3,x\right)={e^{i p_3^2x}\over 2}\left[1-(1+i)C\left(\sqrt{2x\over\pi}p_3\right)-(1-i)S\left(\sqrt{2x\over\pi}p_3\right)\right]$$
with $C(x)$ and $S(x)$ the Fresnel integrals. The contribution of the integral path $s=-iS_E+\epsilon e^{i\theta}$ with $\theta\in[\pi,\pi/2]$ is infinitesimally small over the integrand of Eq.\eqref{SMB} thus can be neglected for the Fourier transformation. Similar argument applies to the integral path $\int_{-i(S_E-\epsilon)}^{-iS_E+\epsilon}\theta(-y_3)$. Finally, the path $\int_{RT}$ can be simply represented as $$\int_{\Gamma_\epsilon}\equiv\int_{0}^{\infty}-\int_{-i(S_E-\epsilon)}^{\infty-i(S_E-\epsilon)}$$
for the quark propagator in energy momentum space, as no singularity is encountered during the deformation of the integral path for $y_3>0$. The aforementioned delta function terms are similar to the last term in the square bracket of Eq.\eqref{SMB}, thus cancel out due to the fact that $\partial_{x_3}\theta(y)=-\partial_{x_3}\theta(-y)$ and the continuity of the integrand around $s=-iS_E$. Now, it is straightforward to take full Fourier transformation of the effective propagator Eq.~\eqref{effprop} to get
\begin{eqnarray}
\hat{\cal S}({p})
&=&i\int_{\Gamma_\epsilon} {\di s}\exp\left\{-i (M^2-i\eta)s-i{\tan(qBs)\over qB}(p_1^2+p_2^2)-i{\tanh(qEs)\over qE}(-{p}_0^2+p_3^2)\right\}\nonumber\\
&&\Big[m\!-\!i~\gamma^5\pi^0+i\,\gamma^4(p_0+{\tanh(qEs)}p_3)\!-\!\gamma^3(p_3+\,{\tanh(qEs)}p_0)\!-\!\gamma^2(p_2+{\tan(qBs)}p_1)\nonumber\\
&&-\gamma^1(p_1-{\tan(qBs)}p_2)\Big]\Big[1+{i\gamma^5\tanh(qEs)\tan(qBs)}
+{\gamma^1\gamma^2\tan(qBs)}-i{\gamma^4\gamma^3\tanh(qEs)}\Big],\label{propm}
\end{eqnarray}
the integrand of which is the same as that in IOF~\cite{Wang:2018gmj}.

As the delta functions in the quark propagator do not play roles in the evaluations of the order parameters, the traces over the propagator can be carried out explicitly to obtain
\begin{eqnarray}
{1\over V_4}\text {Tr}\;{\cal S}(x,x)
	&=&N_cm\,K(M,E,B)-N_c{\pi^0qEqB\over 4\pi^2M^2}\left(1-e^{-M^2S_E}\right),\label{Trm}\\
	{1\over V_4}\text {Tr}\;{\cal S}(x,x)i\gamma^5
		&=&N_c\pi^0\,K(M,E,B)+N_c{mqEqB\over 4\pi^2M^2}\left(1-e^{-M^2S_E}\right),\label{Trpi0}\\
	K(M,E,B)&=&\Re\left\{\int_{0}^{\infty}{\di s\over 4\pi^2}\;e^{-M^2s}\left[{qB\over\tanh(qBs)}{qE\over\tan(qEs)}-e^{-M^2S_E}{qB\over\tanh(q B(s+S_E)}{qE\over\tan(qEs)}\right]\right\},
\end{eqnarray}
where we can easily identify the feedback of SPP from the terms with $e^{-M^2S_E}$. 
Similar to the "vacuum regularization" scheme, we introduce subtract terms that make sure the convergence of the proper-time integral and then are presented by three-momentum cutoff, that is,
\begin{eqnarray}
K^r(M,E,B)&=&\Re\left\{\int_{0}^{\infty}{\di s\over 4\pi^2}\;e^{-M^2s}\left[{qB\over\tanh(qBs)}{qE\over\tan(qEs)}-{1\over s^2}-e^{-M^2S_E}\left({qB\over\tanh(q B(s+S_E)}{qE\over\tan(qEs)}\right.\right.\right.\nonumber\\
&&\left.\left.\left.-{1\over s(s+\tilde{S}_E(1))}\right)\right]\right\}+F_{\Lambda}^{0^+}(M)-e^{-M^2S_E}F_{\Lambda}^{\tilde{S}_E(1)}(M).
\end{eqnarray}
Here, $\tilde{S}_E(x)=\tanh(x\,qBS_E)/qB$ and the auxiliary function is defined as
\begin{eqnarray}
F_{\Lambda}^{S}(M)\equiv \int^{\Lambda}{\di^3 p\over 2\pi^3}\int_{-\infty}^{\infty}{\di p_4\over 2\pi}{e^{-p_\bot^2S}\over p_4^2+E_p^2}=\int_0^{\Lambda}{p\di p\over \pi^{2}}{D\left(pS^{1/2}\right)\over S^{1/2}E_p}
\end{eqnarray}
with $p_\bot^2=p_1^2+p_2^2, E_p=\sqrt{{\bf p}^2+M^2}$ and $D(x)$ the Dawson's integral function. In the large $qB$ limit, $\tilde{S}_E(1)\rightarrow 0^+$ and $K^r(M,E,B)$ can be simply presented as the form $(1-e^{-M^2S_E})\tilde{K}^r(M,E,B)$ with $\tilde{K}^r(M,E,B)$  the same as that in the IOF. Such a factorization is consistent with that given in Ref.~\cite{Copinger:2018ftr}. The explicit forms of the "gap equations" in IIF can then be obtained by substituting Eqs.~\eqref{Trm} and \eqref{Trpi0} into Eqs.~\eqref{gapm} and \eqref{gappi0}, respectively. As long as the coupled "gap equations" can be solved self-consistently, the neutral pion condensate can be found to be simply
$$\pi^0={qEqB\over 4\pi^2}\left(1-e^{-M^2S_E}\right){2G\over m_0},$$
which is different from the one in IOF~\cite{Cao:2015cka} by the extra term from Schwinger mechanism.

As we've already mentioned, there is no self-consistent thermodynamic potential for both the "gap equations" in IIF, because 
$${\partial\over\partial\pi^0}{1\over V_4}\text {Tr}\;{\cal S}(x,x)\neq {\partial\over\partial m}{1\over V_4}\text {Tr}\;{\cal S}(x,x)i\gamma^5$$
due to the SPP contributions in the last anomaly terms of Eqs.\eqref{Trm} and \eqref{Trpi0}. This is not a big deal since the system is non-equilibrium {\it ab initio} in IIF and the thermodynamic potential or pressure might not be well defined. Nevertheless, there is a consistent thermodynamic potential for the anomaly irrelevant terms:
\begin{eqnarray}
\Omega_{\rm IIF}
&=&{(m-m_0)^2+(\pi^0)^2\over 4G}+N_c\,\Re\Bigg\{\int_{0}^{\infty}{\di s\over 8\pi^2}\;{e^{-M^2s}\over s}\left[{qB\over\tanh(qBs)}{qE\over\tan(qEs)}-{1\over s^2}-{s\,e^{-M^2S_E}\over s+S_E}\left({qB\over\tanh(q B(s+S_E)}\right.\right.\nonumber\\
&&\left.\left.{qE\over\tan(qEs)}-{1\over s(s+\tilde{S}_E(1) )}\right)\right]-G_{\Lambda}^{0^+}(M)+G_{\Lambda}^{\tilde{S}_E(1)}(M)\Bigg\},
\end{eqnarray}
where the auxiliary function is defined as
\begin{eqnarray}
G_{\Lambda}^{S}(M)&\equiv&\int_0^M M'\di M'~e^{-{M'}^2S_E}F_{\Lambda}^{S}(M')=\int_0^{\Lambda}{p\di p\over 2\pi^{3/2}}{e^{p^2S_E}\left[{\rm Erf}\left({S}_E^{1/2}E_p\right)-{\rm Erf}\left({S}_E^{1/2}p\right)\right]{D}\left(pS^{1/2}\right)\over {S}_E^{1/2}S^{1/2}}
\end{eqnarray}
with ${\rm Erf}(x)$ the error function. We'd like to point out that: When chiral anomaly is not relevant to the ground state, such as in pure electric field and beyond the chiral rotation region in PEM, the quantity $-\Omega_{IIF}+\Omega_{IIF}|_{E,B\rightarrow0}$ can be roughly considered as the in-in pressure of the system. A part of the pressure is then from the persistent pair production. 

In the end, it is useful to give the thermodynamic potential in IOF here, that is, $\Omega_{\rm IOF}\equiv\bar{\Omega}_{\rm IOF}+\Omega_{\rm a}$~\cite{Wang:2017pje} with the anomaly term $\Omega_{\rm a}=-N_c{qEqB\over 4\pi^2}\theta$ and the regular one
\begin{eqnarray}
\bar{\Omega}_{\rm IOF}={M^2\!-\!2Mm_0\cos\theta\!+\!m_0^2\over 4G}\!+\!N_c\,\Re\left\{\int_{0}^{\infty}\!\!{\di s\over 8\pi^2}{e^{-M^2s}\over s}\left[{qB\over\tanh(qBs)}{qE\over\tan(qEs)}-{1\over s^2}\right]\!-\!G_{\Lambda}^{0^+}(M)\right\},
\end{eqnarray}
\end{widetext}
where the chiral condensates are presented alternatively in terms of the magnitude $M$ and phase $\theta=\arcsin(\pi^0/M)=\arccos(m/M)$. Though the anomaly part $\Omega_{\rm a}$ is a linear term of the angle variable $\theta$, the $2\pi$ periodicity is automatically guaranteed for the physical results if we check the more fundamental quantity: the partition function of the system in real-time, that is
\bea
{\cal{Z}}_{\rm a}=e^{-V_4\Omega_{\rm a}}=e^{i\,Q\theta},\ Q=\left[VN_c{qEqB\over 4\pi^2}\right].
\eea
Here, we defined  the winding number $Q\in\mathbb{N}^+$ for $qEqB>0$ by following the spirit of chiral anomaly~\cite{Weinberg1996} and  the symbol $[x]$ denotes natural number equal to or smaller than positive $x$. For large $Q$ as expected if the time-space volume $V$ is very large, the approximation $Q\approx VN_c{qEqB\over 4\pi^2}$ can be used as long as we constrain ourselves to one period regime of $\theta$ by applying ${\rm mod}(\theta,2\pi)$. However, the preferred value of $\theta$ depends on the regime we choose for ${\rm mod}(\theta,2\pi)$ and is a boundary minimum, which means not any value is really preferred at all and the phase can change randomly. For later convenience, we get rid of the ${\rm mod}$ function here as long as no physical ambiguity is induced. 

\section{Polarization Functions in Real-time Formalism}\label{PFIIF}
As the lengthy derivations of the polarization functions for the neutral sector were already given in Ref.~\cite{Wang:2018gmj} for IOF, we present the corresponding derivations and regularizations for IIF in this section. Adopting the effective quark propagator in Eq.\eqref{propm}, the polarization function of $ \hat{\pi}^0 $ boson with only energy $ q_0 $ nonzero can be evaluated as the following:
\begin{widetext}
	\begin{eqnarray}\label{Pipp}
	&&\Pi_{\hat{\pi}^0\hat{\pi}^0}({B},{E},q_0)\equiv-i\int\frac{d^4 p}{(2\pi)^4}{\rm Tr}   \hat{S}(p+q_0)i\gamma_5\hat{S}(p)i\gamma_5\nonumber\\
	&=&N_c\frac{qEqB}{4\pi^2}\int_{\Gamma_\epsilon}ds\int_{\Gamma_\epsilon}ds'~{\rm exp}\Bigg\{-i \left[(M^2-i\eta)s_+-\frac{\tanh(qEs)\tanh(q Es')}{q E(\tanh(q Es)+\tanh(q Es'))}q_0^2\right]\Bigg\}\ \ \Bigg\{-2m\pi^0+\nonumber\\
	&&\frac{(\pi^0)^2-m^2}{{\rm tanh}(q Es_+){\rm tan}(q Bs_+)}+\frac{i q B {\rm sin^{-2}}(q Bs_+)}{{\rm tanh}(q Es_+)}+\frac{i q E {\rm sinh^{-2}}(q Es_+)}{{\rm tan}(q Bs_+)}-\frac{q_0^2 {\rm tanh}(q Es){\rm tanh}(q Es'){\rm sinh^{-2}}(q Es_+)}{({\rm tanh}(q Es)+{\rm tanh}(q Es')){\rm tan}(q Bs_+)}\Bigg\}\nonumber\\
	&{=}&\frac{N_c}{8\pi^2}\int_{0}^\infty tdt\int_{-1}^1 du~{\rm exp}\Bigg\{-i \left[(M^2-i\eta)t-\frac{\tanh(qEt^+)\tanh(q Et^-)}{q E(\tanh(q Et^+)+\tanh(q Et^-))}q_0^2\right]\Bigg\}\ \sum_{x=0}^2\bar\Pi_{\hat{\pi}^0\hat{\pi}^0}^{x}(E,B,q_0),
	\end{eqnarray}
	where $s_\pm=s\pm s', t^\pm=\frac{t(1\pm u)}{2}$ and the auxiliary functions in the integrand are defined as
	\begin{eqnarray}
	&&\bar\Pi_{\hat{\pi}^0\hat{\pi}^0}^{0}(E,B,q_0)/(qEqB)\nonumber\\
	&=&\Bigg\{-2m\pi^0-\frac{q_0^2 \tanh(q Et^+)\tanh(q Et^-)\sinh^{-2}(q Et)}{(\tanh(q Et^+)+\tanh(q Et^-)){\tan}(q Bt)}+\frac{1}{\tanh(q Et)\tan(q Bt)}\Bigg[\frac{i}{t}+2(\pi^0)^2\nonumber\\
	&&\ \ +\frac{q_0^2}{2} \text{csch}(q Et) \Big(u\sinh (q Etu)-\coth (q Et)\cosh(q Etu)+\text{csch}(q Et)\Big)\Bigg]\Bigg\},\\
	&&\bar\Pi_{\hat{\pi}^0\hat{\pi}^0}^{x}(E,B,q_0)/(qEqB)\nonumber\\
	&=&f(x)e^{-xM^2S_E}\Bigg\{-2m\pi^0+\frac{(\pi^0)^2-m^2}{{\rm tanh}(q Et){\rm tan}(q B(t-i\,x S_E))}+\frac{i q B {\rm sin^{-2}}(q B(t-i\,x S_E))}{{\rm tanh}(q Et)}+\frac{i q E {\rm sinh^{-2}}(q Et)}{{\rm tan}(q B(t-i\,x S_E))}\nonumber\\
	&&-\frac{q_0^2 {\rm tanh}(q Et^+){\rm tanh}(q Et^-){\rm sinh^{-2}}(q Et)}{({\rm tanh}(q Et^+)+{\rm tanh}(q Et^-)){\rm tan}(q B(t-i\,x S_E))}\Bigg\},\ \ x=1,2\label{pipi}
	\end{eqnarray}
	with $f(x)=3x^2-6x+1$. In the last step of Eq.\eqref{Pipp}, we've used partial integral to remove the seemed divergent term $\sim{\rm sin^{-2}}(q Bt)$ in  $\bar\Pi_{\hat{\pi}^0\hat{\pi}^0}^{0}$ according to the non-overlapping condition~\cite{Schwinger1973}, which implies that the higher order divergence at $t=0$ is nonphysical and can be reduced to $\sim{\cot}(q Bt)$ through partial integral. Besides, the divergence at $t=n\pi\ (n\in \mathbb{N}_+)$ is an artifact that  is introduced when we stick to the integration path of $t$ along real axis by taking its infinitesimal negative  imaginary component~\cite{Copinger:2018ftr} to zero. The application of non-overlapping condition avoids the artifact automatically when Cauchy principal value integration is kept in mind.
	
Then, $\bar\Pi_{\hat{\pi}^0\hat{\pi}^0}^{0}$ can be regularized in the same way as that in Ref.~\cite{Wang:2018gmj}. For the terms with $x=1,2$, the integrations can be regularized in the "subtraction-addition" scheme, that has been utilized for the gap equations, as
	\begin{eqnarray}
	&&\Pi_{\hat{\pi}^0\hat{\pi}^0}^{x\,r}(B,E,q_0)\nonumber\\
	&=&\frac{N_c}{8\pi^2}\int_{0}^\infty tdt\int_{-1}^1 du~\Bigg({\rm exp}\Bigg\{-i \left[(M^2-i\eta)t-\frac{\tanh(qEt^+)\tanh(q Et^-)}{q E(\tanh(q Et^+)+\tanh(q Et^-))}q_0^2\right]\Bigg\}\ \bar\Pi_{\hat{\pi}^0\hat{\pi}^0}^{x}(E,B,q_0)\nonumber\\
	&&-f(x)e^{-xM^2S_E}{e^{-i \,t\left[(M^2-i\eta)-{1-u^2\over 4}q_0^2\right]}}\Bigg\{\frac{(\pi^0)^2-m^2}{t(t-i\tilde{S}_E(x))}+\frac{i}{t(t-i \tilde{S}_E(x))^2}+\frac{i}{t^2(t-i\tilde{S}_E(x))}-\frac{q_0^2 \frac{(1-u^2)}{4}}{t(t-i\tilde{S}_E(x))}\Bigg\}\Bigg)\nonumber\\
	&&+\Pi_{\hat{\pi}^0\hat{\pi}^0}^{x\, \Lambda}(E,B,q_0).\label{Pi_pipir}
	\end{eqnarray}
 Altogether, the regularized polarization function for $ \hat{\pi}^0 $ meson is just 
	$$\Pi_{\hat{\pi}^0\hat{\pi}^0}^r({B},{E},q_0)=\sum_{x=0}^2\Pi_{\hat{\pi}^0\hat{\pi}^0}^{x\,r}(B,E,q_0),$$
	and the regularized $\hat\sigma$- and mixing-mode polarization functions can be given similarly as the following:
	\begin{eqnarray}
	&&\Pi_{\hat{\sigma}\hat{\sigma}}^r({B},{E},q_0)\equiv{\rm Reg}\left[-i\int\frac{\di^4 \p}{(2\pi)^4}{\rm Tr}\hat{S}(p+q_0)\hat{S}(p)\right]=\Pi_{\hat{\pi}^0\hat{\pi}^0}^r({B},{E},q_0)|_{m\rightarrow-\pi^0}^{\pi^0\rightarrow m},\\
	&&\Pi_{\hat{\sigma}\hat{\pi}^0}^r({B},{E},q_0)=\Pi_{\hat{\pi}^0\hat{\sigma}}^r({B},{E},q_0)\equiv{\rm Reg}\left[-i\int\frac{\di^4 \p}{(2\pi)^4}{\rm Tr}   \hat{S}(p+q_0)i\gamma_5\hat{S}(p)\right]\nonumber\\
	&=&\frac{N_c}{8\pi^2}\int_{0}^{\infty}\!\!tdt\int_{-1}^{1}\!\!du~\sum_{x=0}^2f(x)e^{-xM^2S_E}\Bigg({\rm exp}\Bigg\{-i\Bigg[(M^2\!-\!i\eta)t-\frac{\tanh(qEt^+)\tanh(q Et^-)}{q E(\tanh(q Et^+)\!+\!\tanh(q Et^-))}q_0^2\Bigg]\Bigg\}~\Bigg[qEqB\nonumber\\
	&&(m^2\!-\!(\pi^0)^2)+\frac{2qEqB\,m\pi^0}{\tan(q B(t-i\,xS_E))\tanh(q Et)}\Bigg]-{e^{-i \,t\left[(M^2-i\eta)-{1-u^2\over 4}q_0^2\right]}}\frac{2m\pi^0}{t(t-i\tilde{S}_E(x))}\Bigg)+\sum_{x=0}^2\Pi_{\hat{\sigma}\hat{\pi}^0}^{x\, \Lambda}(E,B,q_0).\label{Pi_sigmapir}
	\end{eqnarray}
Finally, the "addition" terms  can be evaluated alternatively within the three-momentum cutoff scheme~\cite{Wang:2018gmj} as:
	\bea
	&&\Pi_{\hat{\pi}^0\hat{\pi}^0}^{x\, \Lambda}(E,B,q_0)\nonumber\\
	&=&\frac{N_c}{4\pi^2}f(x)e^{-xM^2S_E}\int_{0}^\infty ds \int_{0}^\infty ds'~\frac{e^{-i \left[(M^2-i\eta)s_+-{ss'\over s_+}q_0^2\right]}}{s_+(s_+-i\,\tilde{S}_E(x))}\left[{(\pi^0)^2-m^2}+\frac{i}{s_+-i\,\tilde{S}_E(x)}+\frac{i}{s_+}-q_0^2{ss'\over s_+^2}\right]\nonumber\\
	&=&i\frac{N_c}{4\pi^4}f(x)\int\di^4\p~e^{-xM^2S_E-\tilde{S}_E(x)p_\bot^2}\int_{0}^\infty ds\int_{0}^\infty ds'~e^{-i \left[(E_p^2-i\eta)s_+-s'p_0^2-s(p_0+q_0)^2\right]}\Big[(\pi^0)^2-m^2-{\bf p}^2+(p_0+q_0)p_0\Big]\nonumber\\
	&=&-i\frac{N_c}{4\pi^4}f(x)\int\di^4\p~e^{-xM^2S_E-\tilde{S}_E(x)p_\bot^2}{(\pi^0)^2\!-\!m^2\!-\!{\bf p}^2\!+\!(p_0\!+\!q_0)p_0 \over [(p_0+q_0)^2-E_p^2][p_0^2-E_p^2]}=\frac{N_c}{\pi^3}f(x)\int\di^3{\bf p}~e^{-xM^2S_E-\tilde{S}_E(x)p_\bot^2}{m^2+{\bf p}^2 \over E_p(q_0^2-4E_p^2)}\nonumber\\
	&=&\frac{4N_c}{\pi^2}f(x)e^{-xM^2S_E}\int_0^\Lambda p^2\di p~{D(\tilde{S}_E^{1/2}(x)p)\over \tilde{S}_E^{1/2}(x)p}{m^2+{p}^2 \over E_p(q_0^2-4E_p^2)},
	\eea
	\bea
   &&\Pi_{\hat{\sigma}\hat{\pi}^0}^{x\, \Lambda}(E,B,q_0)\nonumber\\
	&=&m\pi^0\frac{N_c}{2\pi^2}f(x)e^{-xM^2S_E}\int_{0}^\infty ds\int_{0}^\infty ds'~\frac{e^{-i \left[(M^2-i\eta)s_+-{ss'\over s_+}q_0^2\right]}}{s_+(s_+-i\,\tilde{S}_E(x))}\nonumber\\
	&=&i\,m\pi^0\frac{N_c}{2\pi^4}f(x)\int\di^4\p~e^{-xM^2S_E-\tilde{S}_E(x)p_\bot^2}\int_{0}^\infty ds\int_{0}^\infty ds'~e^{-i \left[(E_p^2-i\eta)s_+-s'p_0^2-s(p_0+q_0)^2\right]}\nonumber\\
	&=&-i\frac{N_c}{2\pi^4}f(x)\int\di^4\p~e^{-xM^2S_E-\tilde{S}_E(x)p_\bot^2}{m\pi^0\over [(p_0+q_0)^2-E_p^2][p_0^2-E_p^2]}=-\frac{N_c}{\pi^3}f(x)\int\di^3{\bf p}~e^{-xM^2S_E-\tilde{S}_E(x)p_\bot^2}{m\pi^0\over E_p(q_0^2-4E_p^2)}\nonumber\\
	&=&-\frac{4N_c}{\pi^2}f(x)e^{-xM^2S_E}\int_0^\Lambda p^2\di p~{D(\tilde{S}_E^{1/2}(x)p)\over \tilde{S}_E^{1/2}(x)p}{m\pi^0\over E_p(q_0^2-4E_p^2)},
	\eea
which are consistent with the regularized vacuum terms $\Pi_{\hat{\pi}^0\hat{\pi}^0}^{\Lambda}$ and $\Pi_{\hat{\sigma}\hat{\pi}^0}^{\Lambda}$ in Ref.~\cite{Wang:2018gmj} if we set $x=0$.
	
\end{widetext}

\section{Numerical calculations}\label{Numer}
In order to perform numerical calculations and compare to QCD, the three parameters of the NJL model are fixed to $G=9.86~{\rm GeV}^{-2}$, $\Lambda=0.653~{\rm GeV}$ and $m_0=5~{\rm MeV}$ by fitting the neutral pion mass $m_\pi=134~{\rm MeV}$, pion decay constant $f_\pi=93~{\rm MeV}$ and chiral condensate $\langle\bar\psi\psi\rangle=-(0.25~{\rm GeV})^3$ physically~\cite{Zhuang:1994dw}. For the purpose of improving the convergence of the proper-time integrals, especially those involved in the polarization functions, the variable $s$ can be rotated by an angle to ${1-C_si\over\sqrt{1+C_s^2}}s~(C_s\geq 0)$ in the complex plane. We've checked carefully that the numerical results do not depend on the value of $C_s$ as no singularity is ever encountered during the rotation, so we simply set $C_s=1$ in the calculations. Another trick that can be adopted for the sake of convergence, especially when $q_0\gtrsim 2M$, is that the subtract terms in Eqs.~\eqref{Pi_pipir} and \eqref{Pi_sigmapir} can be approximated by the ones with small enough EM field while keeping $S_E$ and $\tilde{S}_E(x)$ fixed. Take the simplest term in $\Pi^r_{\hat{\sigma}\hat{\pi}^0}$ for example, we can take the following approximation:
\begin{eqnarray}
&&\frac{2m\pi^0~{e^{-i \,t\left[(M^2-i\eta)-{1-u^2\over 4}q_0^2\right]}}}{t(t-i\tilde{S}_E(x))}\nonumber\\
&\approx&\frac{2q E_0m\pi^0~e^{-i\big[(M^2\!-\!i\eta)t-\frac{\tanh(qE_0t^+)\tanh(q E_0t^-)q_0^2}{q E_0(\tanh(q E_0t^+)+\tanh(q E_0t^-))}\big]}}{\tanh(q E_0t)(t-i\tilde{S}_E(x))}\nonumber
\end{eqnarray}
with $eE_0=10^{-3}~{\rm GeV}^2$.

\subsection{Pure electric field}
\begin{figure}[!htb]
	\centering
	\includegraphics[width=0.41\textwidth]{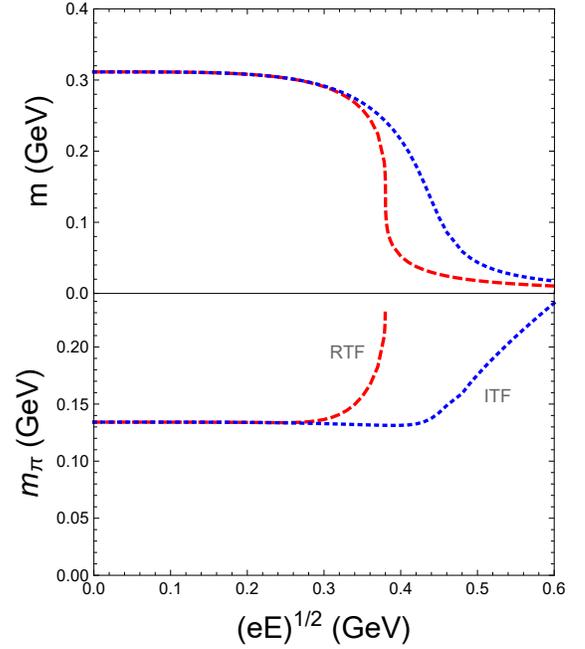}
	\caption{The dynamical quark mass $m$ and neutral pion mass $m_\pi$ as functions of external electric field $eE$ in the IOF (blue dotted lines) and IIF (red dashed lines), respectively.}\label{PEF}
\end{figure}

As a warm-up, we first consider the much simpler pure electric field limit with $B\rightarrow 0$, thus no chiral anomaly is induced by the background EM field and the pseudoscalar condensate $\pi^0=0$~\cite{Cao:2015dya}. Illuminated in the upper and lower panels of Fig.~\ref{PEF} are the evolutions of the order parameter $m$ for chiral symmetry and the pole mass $m_\pi$ of the Goldstone-like mode $\hat{\pi}^0$ with electric field $eE$, respectively.
Similar features are found in the IOF and IIF: With $m$ continuously decreasing with $eE$, the pion mass $m_\pi$ decreases (though slightly) and then increases across the transition region, which is pretty like the crossover feature at finite temperature~\cite{Zhuang:1994dw,Klevansky:1989vi}. However, compared to the results in the IOF, we find that the feedback of SPP becomes prominent around the point $qE\sim m^2$ in the IIF, which then further catalyzes chiral symmetry restoration due to the medium effect. As a consequence, the variations of $m$ and $m_\pi$ become much stiffer around the transition region -- actually, in chiral limit, the chiral symmetry restoration is of second-order in the IOF but of first-order in the IIF. 

\subsection{Parallel electromagnetic field}
\begin{figure}[!htb]
	\centering
	\includegraphics[width=0.41\textwidth]{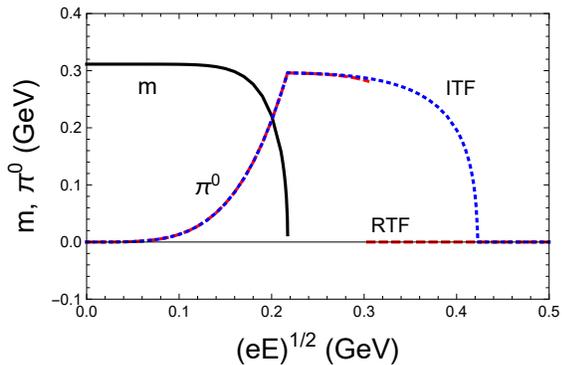}
	\caption{The illuminations of $\pi^0$ as a function of electric field $eE$ in the IOF (blue dotted lines) and IIF (red dashed lines), respectively. The evolutions of $m$ are almost the same up to the end of chiral rotation thus are denoted by a single black solid line.}\label{cond_PEM}
\end{figure}

In the parallel electromagnetic field with $B=E$, the comparisons between the chiral condensates in the IOF and IIF are illuminated in Fig.~\ref{cond_PEM}: The evolutions of $m$ are almost the same up to the end of chiral rotation thus are denoted by a single black solid line. As we can see, the value of $\pi^0$ diverges beyond the point $(qE)^{1/2}\geq \pi^0$ where the SPP  starts to play an important role. As demonstrated clearly by the feature of $m$, our recent more careful check indicates that the phase transition is not of second-order~\cite{Cao:2015cka,Wang:2018gmj} but rather of very weak first-order at the end of chiral rotation with $(eE_{c1})^{1/2}=0.217~{\rm GeV}$. Surprisingly, the chiral restoration becomes of first-order in the IIF rather than second-order in the IOF (see the curves of $\pi^0$). Nevertheless, this is qualitatively consistent with the stiffer feature found in the previous case with pure electric field and the fact that  the Schwinger mechanism is enhanced due to the presence of parallel magnetic field~\cite{Copinger:2016llk}.

\begin{figure}[!htb]
	\centering
	\includegraphics[width=0.41\textwidth]{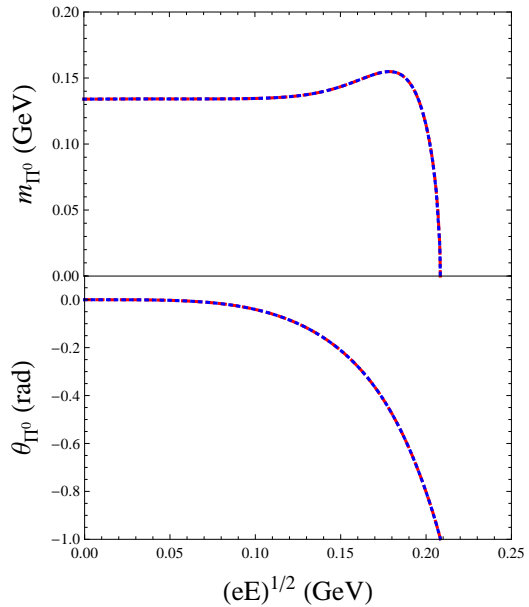}
	\caption{The mass  $m_{\Pi^0}$ of  the Goldstone-like collective mode and the corresponding parity mixing angle $\theta_{\Pi^0}$ as functions of electric field $eE$ in the IOF (blue dotted lines) and IIF (red dashed lines), respectively. $\theta_{\Pi^0}=0$ corresponds to the normal pseudoscalar meson $\pi^0$.}\label{mpi_PEM}
\end{figure}

\begin{figure}[!htb]
	\centering
	\includegraphics[width=0.41\textwidth]{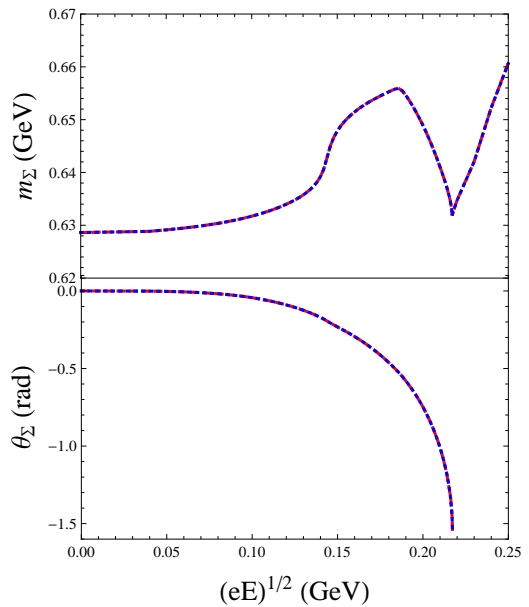}
	\caption{The mass $m_{\Sigma}$ of the Higgs-like collective mode and the corresponding parity mixing angle $\theta_{\Sigma}$ as functions of electric field $eE$ in the IOF (blue dotted lines) and IIF (red dashed lines), respectively. $\theta_{\Sigma}=0$ corresponds to the normal scalar meson $\sigma$.}\label{msigma_PEM}
\end{figure}

Accordingly, the mass eigenstates of the collective modes: the Goldstone-like $\Pi^0$ and the Higgs-like $\Sigma$ are studied and their pole masses are illuminated together with the parity mixing angles in Figs.~\ref{mpi_PEM} and \ref{msigma_PEM}, respectively. Compared to our previous explorations~\cite{Wang:2018gmj}, the recent more precise calculations find nonmonotonic features of both $\Pi^0$ and $\Sigma$ masses with EM field. However, the mixing angles both monotonically decrease from $0$ to $\sim-\pi/2$, which indicates the role exchange between the parity eigenstates $\hat\pi^0$ and $\hat\sigma$, consistent with the condensate exchange during the chiral rotation as shown in Fig.\ref{cond_PEM}. 

\begin{figure}[!htb]
	\centering
	\includegraphics[width=0.41\textwidth]{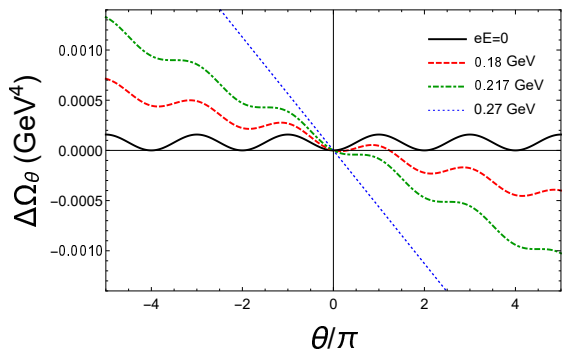}
	\caption{The $\theta$ dependence of the thermodynamic potential $\Delta\Omega_\theta=\Omega_\theta-\Omega_0$ for chosen electric fields $eE=0, 0.18, 0.217$ and $0.27~{\rm GeV}$. The blue dotted line corresponds to the free case with $\Delta\Omega_\theta=-N_c{qEqB\over 4\pi^2}\theta$.}\label{Omega_theta}
\end{figure}

One terrible thing happens around the end point of chiral rotation: the $\Pi^0$ mass decreases to zero, which may imply the instability of the ground state with only $\pi^0$ condensation by following Ginzburg-Landau theory. We plot the $\theta$ dependence of the thermodynamic potential $\Delta\Omega_\theta=\Omega_\theta-\Omega_0$ for chosen EM fields in Fig.~\ref{Omega_theta}. In the vacuum without EM field, there are infinite degenerate global minima locating at $\theta=2k\pi~(k\in \mathbb{Z})$, which is consistent with $\sigma$ condensation found before~\cite{Zhuang:1994dw,Klevansky:1989vi}. With increasing EM field, the global minima become local ones locating at $\theta=\theta_{EM}+2k\pi$ with $\theta_{EM}\in (-\pi,\pi)$ EM-dependent and $\pi^0$ condensate also forms. Eventually, at the critical point $eE_{c1}$, the local minima disappear and $\Delta\Omega_\theta$ becomes a monotonic function of $\theta$, just like the term induced by chiral anomaly: $\Delta\Omega_\theta=-N_c{qEqB\over 4\pi^2}\theta$ in free Fermi gas system with finite fermion mass and electric charge. In the free case, a simple functional variable transformation $\psi\rightarrow e^{-i\gamma^5\theta}\psi$ will give rise to such a term, which then forces the chiral angle $\theta$ to change all the time as there is no minimum. This is the source of instability encountered at $eE_{c1}$: As the expectation value of the chiral phase $\theta$ keeps changing, the phase fluctuation $\Pi^0$ is never a well-defined collective mode. Keep in mind that this is different from the case without EM field and in chiral limit, where the thermodynamic potential $\Omega$ is $\theta$-independent and the massless phase excitation mode itself shifts the expectation value of $\theta$ randomly. 

For the Higgs-like mode, there is no such problem because it corresponds to the magnitude of the "order parameter" of chiral symmetry. As $\theta$ keeps changing, the curves of $\pi^0$ condensate beyond the critical point $eE_{c1}$ in Fig.~\ref{cond_PEM} should be understood as those of the magnitude $M$. Along with the chiral restoration in PEM field, the magnitude excitation mode $\Sigma$ can be studied even beyond $eE_{c1}$: The pole mass of $\Sigma$ starts to increase with EM field, see Fig.~\ref{msigma_PEM}. Thus, even though the chiral restoration feature of $M$ up to $(eE)^{1/2}=0.25~{\rm GeV}$ is quite similar to that at finite temperature, the response of $\Sigma$ mode is rather different. In the previous study,  it was found that~\cite{Klevansky:1989vi}: with temperature increasing up to the pseudo-critical point, the pole mass of $\sigma$ meson decreases down to a value close to that of pion mesons. We've checked that the qualitative difference is not due to the mixing term $\Pi^r_{\hat{\sigma}\hat{\pi}^0}$ induced by PEM field, so the cause might just be the interplay between electric and magnetic field effects; refer to Ref.~\cite{Wang:2017pje} for more discussions on the interplay.

\section{Summary}\label{summary}
In this work, we compare the effects of electromagnetic fields on strong coupling systems within in-out and in-in formalisms under the framework of one-flavor Nambu--Jona-Lasinio model. The main physical difference between the IOF and IIF is that the feedback of Schwinger pair production is also taken into account in the later. In the pure electric field case, we find that the SPP effect further catalyzes chiral symmetry restoration beyond the pair instable point $qE\sim m^2$. This feature remains in parallel EM field, but the effect on chiral rotation is negligible because only small EM field is relevant at this stage. We study the chiral rotation and the corresponding mass eigen collective modes with more precision than that in Ref.~\cite{Wang:2018gmj} and find: the transition is of weak first-order rather than of second-order~\cite{Cao:2015cka} at the end of chiral rotation $eE_{c1}$ and the masses of both the Goldstone-like and Higgs-like modes show nonmonotonic behaviors there.

At the chiral rotation stage, one serious problem is that the mass of the Goldstone-like mode $\Pi^0$ decreases to zero around $eE_{c1}$, which seems to indicate a "wrong" ground state we've proposed. This instability is actually related to the chiral anomaly induced by PEM field and corresponds to a "chaotic" state for the chiral phase, similar to that in free Fermi gas system with finite fermion mass and electric charge. After clarifying the instability, we're now at a point ready to extend the research on PEM effect to the realistic three-flavor case, where more pseudoscalar condensations and parity mixing are involved. 

\emph{Acknowledgments}---
G.C. thanks X.-G. Huang and L. Wang for their comments on this work. G.C. is supported by the National Natural Science Foundation of China with Grant No. 11805290 and Young Teachers Training Program of Sun Yat-Sen University with Grant No. 19lgpy282.

\appendix


\begin{thebibliography}{99}
	
%\cite{Skokov:2009qp}
\bibitem{Skokov:2009qp}
V.~Skokov, A.~Y.~Illarionov and V.~Toneev,
%``Estimate of the magnetic field strength in heavy-ion collisions,''
Int.\ J.\ Mod.\ Phys.\ A {\bf 24}, 5925 (2009).
%doi:10.1142/S0217751X09047570
%[arXiv:0907.1396 [nucl-th]].
%%CITATION =doi:10.1142/S0217751X09047570;%%
%505 citations counted in INSPIRE as of 02 Dec 2017



%\cite{Deng:2012pc}
\bibitem{Deng:2012pc}
W.~T.~Deng and X.~G.~Huang,
%``Event-by-event generation of electromagnetic fields in heavy-ion collisions,''
Phys.\ Rev.\ C {\bf 85}, 044907 (2012).
%doi:10.1103/PhysRevC.85.044907
%[arXiv:1201.5108 [nucl-th]].
%%CITATION =doi:10.1103/PhysRevC.85.044907;%%
%235 citations counted in INSPIRE as of 02 Dec 2017



%\cite{Deng:2014uja}
\bibitem{Deng:2014uja}
W.~T.~Deng and X.~G.~Huang,
%``Electric fields and chiral magnetic effect in Cu+Au collisions,''
Phys.\ Lett.\ B {\bf 742}, 296 (2015).
%doi:10.1016/j.physletb.2015.01.050
%[arXiv:1411.2733 [nucl-th]].
%%CITATION =doi:10.1016/j.physletb.2015.01.050;%%
%24 citations counted in INSPIRE as of 02 Dec 2017

\bibitem{Bloczynski:2012en}
J.~Bloczynski, X.~G.~Huang, X.~Zhang and J.~Liao,
%``Azimuthally fluctuating magnetic field and its impacts on observables in heavy-ion collisions,''
Phys.\ Lett.\ B {\bf 718}, 1529 (2013).
%doi:10.1016/j.physletb.2012.12.030
%[arXiv:1209.6594 [nucl-th]].
%%CITATION =doi:10.1016/j.physletb.2012.12.030;%%
%66 citations counted in INSPIRE as of 02 Dec 2017

%\cite{Gusynin:1994re}
\bibitem{Gusynin:1994re}
V.~P.~Gusynin, V.~A.~Miransky and I.~A.~Shovkovy,
%``Catalysis of dynamical flavor symmetry breaking by a magnetic field in (2+1)-dimensions,''
Phys.\ Rev.\ Lett.\  {\bf 73}, 3499 (1994)
Erratum: [Phys.\ Rev.\ Lett.\  {\bf 76}, 1005 (1996)].
%doi:10.1103/PhysRevLett.73.3499
%[hep-ph/9405262].
%%CITATION =doi:10.1103/PhysRevLett.73.3499;%%
%398 citations counted in INSPIRE as of 02 Dec 2017

%\cite{Gusynin:1994xp}
\bibitem{Gusynin:1994xp} 
V.~P.~Gusynin, V.~A.~Miransky and I.~A.~Shovkovy,
%``Dimensional reduction and dynamical chiral symmetry breaking by a magnetic field in (3+1)-dimensions,''
Phys.\ Lett.\ B {\bf 349}, 477 (1995).
%doi:10.1016/0370-2693(95)00232-A
%[hep-ph/9412257].
%%CITATION = doi:10.1016/0370-2693(95)00232-A;%%
%288 citations counted in INSPIRE as of 21 Nov 2019

%\cite{Bali:2011qj}
\bibitem{Bali:2011qj}
G.~S.~Bali, F.~Bruckmann, G.~Endrodi, Z.~Fodor, S.~D.~Katz, S.~Krieg, A.~Schafer and K.~K.~Szabo,
%``The QCD phase diagram for external magnetic fields,''
JHEP {\bf 1202}, 044 (2012).
%doi:10.1007/JHEP02(2012)044
%[arXiv:1111.4956 [hep-lat]].
%%CITATION =doi:10.1007/JHEP02(2012)044;%%
%352 citations counted in INSPIRE as of 02 Dec 2017



%\cite{Bali:2012zg}
\bibitem{Bali:2012zg}
G.~S.~Bali, F.~Bruckmann, G.~Endrodi, Z.~Fodor, S.~D.~Katz and A.~Schafer,
%``QCD quark condensate in external magnetic fields,''
Phys.\ Rev.\ D {\bf 86}, 071502 (2012).
%doi:10.1103/PhysRevD.86.071502
%[arXiv:1206.4205 [hep-lat]].
%%CITATION =doi:10.1103/PhysRevD.86.071502;%%
%244 citations counted in INSPIRE as of 02 Dec 2017


%\cite{Bruckmann:2013oba}
\bibitem{Bruckmann:2013oba}
F.~Bruckmann, G.~Endrodi and T.~G.~Kovacs,
%``Inverse magnetic catalysis and the Polyakov loop,''
JHEP {\bf 1304}, 112 (2013).
%doi:10.1007/JHEP04(2013)112
%[arXiv:1303.3972 [hep-lat]].
%%CITATION =doi:10.1007/JHEP04(2013)112;%%
%157 citations counted in INSPIRE as of 02 Dec 2017

%\cite{Fukushima:2012kc,Kojo:2012js,Chao:2013qpa,Cao:2014uva,Mao:2016fha}
\bibitem{Fukushima:2012kc}
K.~Fukushima and Y.~Hidaka,
%``Magnetic Catalysis Versus Magnetic Inhibition,''
Phys.\ Rev.\ Lett.\  {\bf 110}, no. 3, 031601 (2013).%[arXiv:1209.1319 [hep-ph]].
%%CITATION = ARXIV:1209.1319;%%
%69 citations counted in INSPIRE as of 11 Oct 2015

%\cite{Kojo:2012js}
\bibitem{Kojo:2012js}
T.~Kojo and N.~Su,
%``The quark mass gap in a magnetic field,''
Phys.\ Lett.\ B {\bf 720}, 192 (2013). %[arXiv:1211.7318 [hep-ph]].
%%CITATION = ARXIV:1211.7318;%%
%42 citations counted in INSPIRE as of 11 Oct 2015

%\cite{Hattori:2015aki}
\bibitem{Hattori:2015aki}
K.~Hattori, T.~Kojo and N.~Su,
%``Mesons in strong magnetic fields: (I) General analyses,''
Nucl.\ Phys.\ A {\bf 951} (2016) 1.
%  doi:10.1016/j.nuclphysa.2016.03.016
% [arXiv:1512.07361 [hep-ph]].
%%CITATION = doi:10.1016/j.nuclphysa.2016.03.016;%%
%25 citations counted in INSPIRE as of 16 Nov 2019

%\cite{Chao:2013qpa}
\bibitem{Chao:2013qpa}
J.~Chao, P.~Chu and M.~Huang,
%``Inverse magnetic catalysis induced by sphalerons,''
Phys.\ Rev.\ D {\bf 88}, 054009 (2013).%[arXiv:1305.1100 [hep-ph]].
%%CITATION = ARXIV:1305.1100;%%
%36 citations counted in INSPIRE as of 11 Oct 2015

%\cite{Cao:2014uva}
\bibitem{Cao:2014uva}
G.~Cao, L.~He and P.~Zhuang,
%``Collective modes and Kosterlitz-Thouless transition in a magnetic field in the planar Nambu-Jona-Lasino model,''
Phys.\ Rev.\ D {\bf 90}, no. 5, 056005 (2014).%[arXiv:1408.5364 [hep-ph]].
%%CITATION = ARXIV:1408.5364;%%
%4 citations counted in INSPIRE as of 11 Oct 2015

%\cite{Ferrer:2014qka}
\bibitem{Ferrer:2014qka}
E.~J.~Ferrer, V.~de la Incera and X.~J.~Wen,
%``Quark Antiscreening at Strong Magnetic Field and Inverse Magnetic Catalysis,''
Phys.\ Rev.\ D {\bf 91}, no. 5, 054006 (2015).%[arXiv:1407.3503 [nucl-th]].
%%CITATION = ARXIV:1407.3503;%%
%19 citations counted in INSPIRE as of 11 Oct 2015

%\cite{Mao:2016fha}
\bibitem{Mao:2016fha} 
S.~Mao,
%``Inverse magnetic catalysis in Nambu?Jona-Lasinio model beyond mean field,''
Phys.\ Lett.\ B {\bf 758}, 195 (2016).%doi:10.1016/j.physletb.2016.05.018
%[arXiv:1602.06503 [hep-ph]].
%%CITATION = doi:10.1016/j.physletb.2016.05.018;%%
%18 citations counted in INSPIRE as of 06 Aug 2018

%\cite{Hidaka:2012mz}
\bibitem{Hidaka:2012mz} 
Y.~Hidaka and A.~Yamamoto,
%``Charged vector mesons in a strong magnetic field,''
Phys.\ Rev.\ D {\bf 87}, no. 9, 094502 (2013).%doi:10.1103/PhysRevD.87.094502
%[arXiv:1209.0007 [hep-ph]].
%%CITATION = doi:10.1103/PhysRevD.87.094502;%%
%81 citations counted in INSPIRE as of 25 May 2019

%\cite{Bali:2017ian}
\bibitem{Bali:2017ian} 
G.~S.~Bali, B.~B.~Brandt, G.~Endrodi and B.~Glassle,
%``Meson masses in electromagnetic fields with Wilson fermions,''
Phys.\ Rev.\ D {\bf 97}, no. 3, 034505 (2018).%doi:10.1103/PhysRevD.97.034505
%[arXiv:1707.05600 [hep-lat]].
%%CITATION = doi:10.1103/PhysRevD.97.034505;%%
%24 citations counted in INSPIRE as of 25 May 2019

%\cite{Avancini:2016fgq}
\bibitem{Avancini:2016fgq} 
S.~S.~Avancini, R.~L.~S.~Farias, M.~Benghi Pinto, W.~R.~Tavares and V.~S.~Timóteo,
%``$\pi_0$ pole mass calculation in a strong magnetic field and lattice constraints,''
Phys.\ Lett.\ B {\bf 767}, 247 (2017).
%doi:10.1016/j.physletb.2017.02.002
%[arXiv:1606.05754 [hep-ph]].
%%CITATION = doi:10.1016/j.physletb.2017.02.002;%%
%26 citations counted in INSPIRE as of 06 Jun 2019

%\cite{Wang:2017vtn}
\bibitem{Wang:2017vtn} 
Z.~Wang and P.~Zhuang,
%``Meson properties in magnetized quark matter,''
Phys.\ Rev.\ D {\bf 97}, no. 3, 034026 (2018).
%doi:10.1103/PhysRevD.97.034026
%[arXiv:1712.00554 [hep-ph]].
%%CITATION = doi:10.1103/PhysRevD.97.034026;%%
%6 citations counted in INSPIRE as of 06 Aug 2019

%\cite{Mao:2018dqe}
\bibitem{Mao:2018dqe} 
S.~Mao,
%``Pions in magnetic field at finite temperature,''
Phys.\ Rev.\ D {\bf 99}, no. 5, 056005 (2019).%doi:10.1103/PhysRevD.99.056005
%[arXiv:1808.10242 [nucl-th]].
%%CITATION = doi:10.1103/PhysRevD.99.056005;%%
%1 citations counted in INSPIRE as of 25 May 2019

%\cite{Liu:2018zag}
\bibitem{Liu:2018zag} 
H.~Liu, X.~Wang, L.~Yu and M.~Huang,
%``Neutral and charged scalar mesons, pseudoscalar mesons, and diquarks in magnetic fields,''
Phys.\ Rev.\ D {\bf 97}, no. 7, 076008 (2018).%doi:10.1103/PhysRevD.97.076008
%[arXiv:1801.02174 [hep-ph]].
%%CITATION = doi:10.1103/PhysRevD.97.076008;%%
%7 citations counted in INSPIRE as of 25 May 2019

%\cite{Coppola:2018vkw}
\bibitem{Coppola:2018vkw} 
M.~Coppola, D.~Gómez Dumm and N.~N.~Scoccola,
%``Charged pion masses under strong magnetic fields in the NJL model,''
Phys.\ Lett.\ B {\bf 782}, 155 (2018).
%doi:10.1016/j.physletb.2018.04.043
%[arXiv:1802.08041 [hep-ph]].
%%CITATION = doi:10.1016/j.physletb.2018.04.043;%%
%12 citations counted in INSPIRE as of 08 Oct 2019
%\cite{Shushpanov:1997sf}

%\cite{Cao:2019res}
\bibitem{Cao:2019res} 
G.~Cao,
%``Magnetic catalysis effect prevents vacuum superconductivity in strong magnetic fields,''
Phys.\ Rev.\ D {\bf 100}, no. 7, 074024 (2019).
%doi:10.1103/PhysRevD.100.074024
%[arXiv:1906.01398 [nucl-th]].
%%CITATION = doi:10.1103/PhysRevD.100.074024;%%
%1 citations counted in INSPIRE as of 25 Nov 2019

%\cite{Zhuang:1994dw}
\bibitem{Zhuang:1994dw}
P.~Zhuang, J.~Hufner and S.~P.~Klevansky,
%``Thermodynamics of a quark - meson plasma in the Nambu-Jona-Lasinio model,''
Nucl.\ Phys.\ A {\bf 576}, 525 (1994).
%doi:10.1016/0375-9474(94)90743-9
%%CITATION =doi:10.1016/0375-9474(94)90743-9;%%
%142 citations counted in INSPIRE as of 02 Dec 2017

%\cite{Klevansky:1989vi}
\bibitem{Klevansky:1989vi}
S.~P.~Klevansky and R.~H.~Lemmer,
%``Chiral symmetry restoration in the Nambu-Jona-Lasinio model with a constant electromagnetic field,''
Phys.\ Rev.\ D {\bf 39}, 3478 (1989).
%doi:10.1103/PhysRevD.39.3478
%%CITATION =doi:10.1103/PhysRevD.39.3478;%%
%212 citations counted in INSPIRE as of 02 Dec 2017

%\cite{Kharzeev:2007jp}
\bibitem{Kharzeev:2007jp}
D.~E.~Kharzeev, L.~D.~McLerran and H.~J.~Warringa,
%``The Effects of topological charge change in heavy ion collisions: 'Event by event P and CP violation',''
Nucl.\ Phys.\ A {\bf 803}, 227 (2008).
%doi:10.1016/j.nuclphysa.2008.02.298
%[arXiv:0711.0950 [hep-ph]].
%%CITATION =doi:10.1016/j.nuclphysa.2008.02.298;%%
%966 citations counted in INSPIRE as of 02 Dec 2017



%\cite{Fukushima:2008xe}
\bibitem{Fukushima:2008xe}
K.~Fukushima, D.~E.~Kharzeev and H.~J.~Warringa,
%``The Chiral Magnetic Effect,''
Phys.\ Rev.\ D {\bf 78}, 074033 (2008).
%doi:10.1103/PhysRevD.78.074033
%[arXiv:0808.3382 [hep-ph]].
%%CITATION =doi:10.1103/PhysRevD.78.074033;%%
%918 citations counted in INSPIRE as of 02 Dec 2017



%\cite{Kharzeev:2010gd}
\bibitem{Kharzeev:2010gd}
D.~E.~Kharzeev and H.~U.~Yee,
%``Chiral Magnetic Wave,''
Phys.\ Rev.\ D {\bf 83}, 085007 (2011).
%doi:10.1103/PhysRevD.83.085007
%[arXiv:1012.6026 [hep-th]].
%%CITATION =doi:10.1103/PhysRevD.83.085007;%%
%183 citations counted in INSPIRE as of 02 Dec 2017



%\cite{Son:2004tq}
\bibitem{Son:2004tq}
D.~T.~Son and A.~R.~Zhitnitsky,
%``Quantum anomalies in dense matter,''
Phys.\ Rev.\ D {\bf 70}, 074018 (2004).
%doi:10.1103/PhysRevD.70.074018
%[hep-ph/0405216].
%%CITATION =doi:10.1103/PhysRevD.70.074018;%%
%193 citations counted in INSPIRE as of 02 Dec 2017



%\cite{Metlitski:2005pr}
\bibitem{Metlitski:2005pr}
M.~A.~Metlitski and A.~R.~Zhitnitsky,
%``Anomalous axion interactions and topological currents in dense matter,''
Phys.\ Rev.\ D {\bf 72}, 045011 (2005).
%doi:10.1103/PhysRevD.72.045011
%[hep-ph/0505072].
%%CITATION =doi:10.1103/PhysRevD.72.045011;%%
%213 citations counted in INSPIRE as of 02 Dec 2017



%\cite{Huang:2013iia}
\bibitem{Huang:2013iia}
X.~G.~Huang and J.~Liao,
%``Axial Current Generation from Electric Field: Chiral Electric Separation Effect,''
Phys.\ Rev.\ Lett.\  {\bf 110}, no. 23, 232302 (2013).
%doi:10.1103/PhysRevLett.110.232302
%[arXiv:1303.7192 [nucl-th]].
%%CITATION =doi:10.1103/PhysRevLett.110.232302;%%
%55 citations counted in INSPIRE as of 02 Dec 2017


%\cite{Hattori:2016njk}
\bibitem{Hattori:2016njk}
K.~Hattori and Y.~Yin,
%``Charge redistribution from anomalous magnetovorticity coupling,''
Phys.\ Rev.\ Lett.\  {\bf 117}, no. 15, 152002 (2016).
%doi:10.1103/PhysRevLett.117.152002
%[arXiv:1607.01513 [hep-th]].
%%CITATION =doi:10.1103/PhysRevLett.117.152002;%%

%\cite{Liao:2014ava}
\bibitem{Liao:2014ava}
J.~Liao,
%``Anomalous transport effects and possible environmental symmetry ¡®violation¡¯ in heavy-ion collisions,''
Pramana {\bf 84}, no. 5, 901 (2015).
%doi:10.1007/s12043-015-0984-x
%[arXiv:1401.2500 [hep-ph]].
%%CITATION =doi:10.1007/s12043-015-0984-x;%%
%52 citations counted in INSPIRE as of 02 Dec 2017



%\cite{Kharzeev:2015znc}
\bibitem{Kharzeev:2015znc}
D.~E.~Kharzeev, J.~Liao, S.~A.~Voloshin and G.~Wang,
%``Chiral magnetic and vortical effects in high-energy nuclear collisions¡ªA status report,''
Prog.\ Part.\ Nucl.\ Phys.\  {\bf 88}, 1 (2016).
%doi:10.1016/j.ppnp.2016.01.001
%[arXiv:1511.04050 [hep-ph]].
%%CITATION =doi:10.1016/j.ppnp.2016.01.001;%%
%155 citations counted in INSPIRE as of 02 Dec 2017



%\cite{Huang:2015oca}
\bibitem{Huang:2015oca}
X.~G.~Huang,
%``Electromagnetic fields and anomalous transports in heavy-ion collisions --- A pedagogical review,''
Rept.\ Prog.\ Phys.\  {\bf 79}, no. 7, 076302 (2016).
%doi:10.1088/0034-4885/79/7/076302
%[arXiv:1509.04073 [nucl-th]].
%%CITATION =doi:10.1088/0034-4885/79/7/076302;%%
%66 citations counted in INSPIRE as of 02 Dec 2017

%\cite{Cao:2015cka}
\bibitem{Cao:2015cka}
G.~Cao and X.~G.~Huang,
%``Electromagnetic triangle anomaly and neutral pion condensation in QCD vacuum,''
Phys.\ Lett.\ B {\bf 757}, 1 (2016).
%doi:10.1016/j.physletb.2016.03.066
%[arXiv:1509.06222 [hep-ph]].
%%CITATION =doi:10.1016/j.physletb.2016.03.066;%%
%12 citations counted in INSPIRE as of 02 Dec 2017

%\cite{Wang:2018gmj}
\bibitem{Wang:2018gmj} 
L.~Wang, G.~Cao, X.~G.~Huang and P.~Zhuang,
%``Nambu?Jona-Lasinio model in a parallel electromagnetic field,''
Phys.\ Lett.\ B {\bf 780}, 273 (2018).
%doi:10.1016/j.physletb.2018.03.018
%[arXiv:1801.01682 [nucl-th]].
%%CITATION = doi:10.1016/j.physletb.2018.03.018;%%
%3 citations counted in INSPIRE as of 29 Sep 2019	

%\cite{Copinger:2016llk}
\bibitem{Copinger:2016llk} 
P.~Copinger and K.~Fukushima,
%``Spatially Assisted Schwinger Mechanism and Magnetic Catalysis,''
Phys.\ Rev.\ Lett.\  {\bf 117}, no. 8, 081603 (2016)
Erratum: [Phys.\ Rev.\ Lett.\  {\bf 118}, no. 9, 099903 (2017)].
%doi:10.1103/PhysRevLett.117.081603, 10.1103/PhysRevLett.118.099903
%[arXiv:1605.05957 [hep-th]].
%%CITATION = doi:10.1103/PhysRevLett.117.081603, 10.1103/PhysRevLett.118.099903;%%
%7 citations counted in INSPIRE as of 25 Nov 2019

%\cite{Copinger:2018ftr}
\bibitem{Copinger:2018ftr} 
P.~Copinger, K.~Fukushima and S.~Pu,
%``Axial Ward identity and the Schwinger mechanism -- Applications to the in-in chiral magnetic effect and condensates,''
Phys.\ Rev.\ Lett.\  {\bf 121}, no. 26, 261602 (2018).
%doi:10.1103/PhysRevLett.121.261602
%[arXiv:1807.04416 [hep-th]].
%%CITATION = doi:10.1103/PhysRevLett.121.261602;%%
%8 citations counted in INSPIRE as of 10 Sep 2019	

%\cite{Cohen:2008wz}
\bibitem{Cohen:2008wz} 
  T.~D.~Cohen and D.~A.~McGady,
  %``The Schwinger mechanism revisited,''
  Phys.\ Rev.\ D {\bf 78}, 036008 (2008).
  
\bibitem{Fradkin:1991} 
E. Fradkin, D. Guitman, and S. Shvartsman, {\it Quantum Electrodynamics: With Unstable Vacuum}, Springer Series
in Nuclear and Particle Physics (Springer-Verlag, Berlin,
1991).	
	
%\cite{Nambu:1961fr}
\bibitem{Nambu:1961fr}
Y.~Nambu and G.~Jona-Lasinio,
%``Dynamical Model Of Elementary Particles Based On An Analogy With Superconductivity. Ii,''
Phys.\ Rev.\  {\bf 124}, 246 (1961).
%doi:10.1103/PhysRev.124.246
%%CITATION =doi:10.1103/PhysRev.124.246;%%
%2413 citations counted in INSPIRE as of 02 Dec 2017



%\cite{Nambu:1961tp}
\bibitem{Nambu:1961tp}
Y.~Nambu and G.~Jona-Lasinio,
%``Dynamical Model of Elementary Particles Based on an Analogy with Superconductivity. 1.,''
Phys.\ Rev.\  {\bf 122}, 345 (1961).
%doi:10.1103/PhysRev.122.345
%%CITATION =doi:10.1103/PhysRev.122.345;%%
%4880 citations counted in INSPIRE as of 02 Dec 2017

%\cite{Wang:2017pje}
\bibitem{Wang:2017pje} 
L.~Wang and G.~Cao,
%``Competition between magnetic catalysis effect and chiral rotation effect,''
Phys.\ Rev.\ D {\bf 97}, no. 3, 034014 (2018).
%doi:10.1103/PhysRevD.97.034014
%[arXiv:1712.09780 [nucl-th]].
%%CITATION = doi:10.1103/PhysRevD.97.034014;%%
%2 citations counted in INSPIRE as of 18 Nov 2019

\bibitem{Weinberg1996}
S. Weinberg, {\it The Quantum Theory of Fields: Morden Applications},  (Cambridge University Press, Cambridge, 1996) Vol. II, Chap. 23, P455-462.

\bibitem{Schwinger1973}
J.~Schwinger, {\it Particles, Sources and Fields} (Addison-Wesley, Reading, Mass. , 1973), Vol. II,
Chap. 4,- Secs. 14-16.
	

%\cite{Cao:2015dya}
\bibitem{Cao:2015dya}
  G.~Cao and X.~G.~Huang,
  %``Chiral phase transition and Schwinger mechanism in a pure electric field,''
  Phys.\ Rev.\ D {\bf 93}, no. 1, 016007 (2016).
  %doi:10.1103/PhysRevD.93.016007
  %[arXiv:1510.05125 [nucl-th]].
  %%CITATION =doi:10.1103/PhysRevD.93.016007;%%
  %4 citations counted in INSPIRE as of 02 Dec 2017

\end{thebibliography}
\end{document}